\documentclass[aps,prl,twocolumn,superscriptaddress,sort&compress]{revtex4-1}%
\usepackage[T1]{fontenc}
\usepackage{lmodern}
\usepackage{eurosym}
\usepackage{amsfonts}
\usepackage{amssymb}
\usepackage{amsmath}
\usepackage{floatrow}
\usepackage[position=t,singlelinecheck=off,caption=false]{subfig}
\usepackage{graphicx}
\usepackage{xcolor}%
\usepackage{booktabs}
\usepackage{adjustbox}
\usepackage[hyperfigures=true,bookmarksnumbered=true,bookmarksopen=true,bookmarks=true,linkcolor=blue,urlcolor=blue,citecolor=blue,colorlinks=true,pdfhighlight=/O,pdfstartview={XYZ null null 1}]{hyperref}
\graphicspath{{}{images/}}
\providecommand{\U}[1]{\protect\rule{.1in}{.1in}}
\providecommand{\U}[1]{\protect\rule{.1in}{.1in}}
\def\HG#1 {\emph{\color{blue}#1}}

\begin{document}
\title{Quantum oscillations and Dirac dispersion in the BaZnBi$_2$ semimetal guaranteed by local Zn vacancy order}
\author{K. Zhao}
\affiliation{Experimentalphysik VI, Center for Electronic Correlations and Magnetism, University of Augsburg, 86159 Augsburg, Germany}
\author{E. Golias}
\affiliation{Helmholtz-Zentrum Berlin für Materialien und Energie, Elektronenspeicherring BESSY II, Albert-Einstein-Stra$\beta$e 15, 12489 Berlin, Germany}
\affiliation{Institut für Experimentalphysik, Freie Universität Berlin, Arnimallee 14, 14195 Berlin, Germany}
\author{Q. H. Zhang}
\affiliation{Beijing National Laboratory for Condensed Matter Physics, Institute of Physics, Chinese Academy of Sciences, Beijing 100190, China}
\author{M. Krivenkov}
\affiliation{Helmholtz-Zentrum Berlin für Materialien und Energie, Elektronenspeicherring BESSY II, Albert-Einstein-Stra$\beta$e 15, 12489 Berlin, Germany}
\author{A. Jesche}
\affiliation{Experimentalphysik VI, Center for Electronic Correlations and Magnetism, University of Augsburg, 86159 Augsburg, Germany}
\author{L. Gu}
\affiliation{Beijing National Laboratory for Condensed Matter Physics, Institute of Physics, Chinese Academy of Sciences, Beijing 100190, China}
\author{O. Rader}
\affiliation{Helmholtz-Zentrum Berlin für Materialien und Energie, Elektronenspeicherring BESSY II, Albert-Einstein-Stra$\beta$e 15, 12489 Berlin, Germany}
\author{I. I. Mazin}
\affiliation{Naval Research Laboratory, code 6390, 4555 Overlook Avenue SW, Washington DC 20375, USA}
\author{P. Gegenwart}
\affiliation{Experimentalphysik VI, Center for Electronic Correlations and Magnetism, University of Augsburg, 86159 Augsburg, Germany}

\begin{abstract}
We have synthesized single crystals of Dirac semimetal candidates AZnBi$_2$ with A=Ba and Sr. In contrast to A=Sr, the Ba material displays a novel local Zn vacancy ordering, which makes the observation of quantum oscillations in out-of-plane magnetic fields possible. As a new Dirac semimetal candidate, BaZnBi$_2$ exhibits small cyclotron electron mass, high quantum mobility, and non-trivial Berry phases. Three Dirac dispersions are observed by ARPES and identified by first-principles band-structure calculations. Compared to AMn(Bi/Sb)$_2$ systems which host Mn magnetic moments, BaZnBi$_2$ acts as non-magnetic analogue to investigate the intrinsic properties of Dirac fermions in this structure family.
\end{abstract}
\maketitle

Topological materials, such as Dirac and Weyl semimetals, have received increasing attention in recent years due to their special physical properties and potential applications~\cite{Na3Bi2012PRB, Cd3As22013PRB, Cd3As22014NM, Na3Bi2014Science, Cd3As2Crystal, Cd3As22014NC, Cd3As22015NM, WeylPRXDFT, WeylPRX, WeylScience, GrapheneNature, TIRMP}. In such systems, the electron transport is often dominated by linearly-dispersing Dirac or Weyl fermions, leading to quantum oscillations with nontrivial Berry phases, quantum Hall effect, and negative magnetoresistance~\cite{Na3Bi2012PRB, Cd3As22013PRB, Cd3As22014NM, Na3Bi2014Science, Cd3As2Crystal, Cd3As22014NC, Cd3As22015NM, WeylPRXDFT, WeylPRX, WeylScience, GrapheneNature, TIRMP}.

Recently, the layered manganese pnictides AMn(Bi/Sb)$_2$ (A=Ca, Sr, Ba, Eu, and Yb) were reported to host Dirac or Weyl fermions, related to their two-dimensional (2D) Bi/Sb layers~\cite{Sr112PRL, Sr112PRB, Ca112PRB, Ca112APL, Sr112DFT, Sr112Neutron, Sr112APRES-PRB, Sr112APRES-SR, Eu112SA, BaMnSb2SR, Ba112PRB, Yb112PRB, SrMnSb2NM, Yb112ARPES, BaMnSb2PNAS, CaMnSb2PRB, Yb112caxis, YbMnSb2PRB}. A common structural feature for all of these compounds are magnetic MnBi/Sb layers intercalated by 2D Bi/Sb layers. The latter have been shown to host quasi-2D Dirac fermions, similar as in graphene and topological insulators~\cite{GrapheneNature, TIRMP, Sr112PRL}. Moreover, the interplay between magnetism and the Dirac dispersion induces novel quantum states: EuMnBi$_2$ shows a bulk half-integer quantum Hall effect due to magnetically confined 2D Dirac fermions~\cite{Eu112SA}. A canted antiferromagnetic order with a ferromagnetic component coexists with the Dirac fermion behavior in Sr$_{1-y}$Mn$_{1-z}$Sb$_2$~\cite{SrMnSb2NM}. Moreover, YbMnBi$_2$ probably features a time-reversal symmetry breaking Weyl fermion state, because the spin degeneracy is lifted by a similar ferromagnetic component~\cite{Yb112ARPES}.

Up to now, all studied compounds of this structure family contain both magnetic MnBi/Sb layers and 2D Bi/Sb layers. To achieve a full understanding of the interplay between the magnetism and Dirac dispersion\cite{Sr112Raman, Sr112Neutron2}, it is instructive to investigate materials without magnetic Mn. This addresses the question, whether a two-dimensional Bi/Sb layer without magnetic layer, still holds Dirac fermions, and if so, to compare its properties with those of the magnetic materials.

Based on powder samples, BaZnBi$_2$ and SrZnBi$_2$ were reported to crystallize in the same tetragonal structure as SrMnBi$_2$~\cite{BaZnBi2}. Below, we report a single crystal study of both materials. While SrZnBi$_2$ shows the expected tetragonal structure with the space group I4/mmm, BaZnBi$_2$ forms local Zn vacancies, which order along a face diagonal direction. This ordering has important consequences regarding the physical properties of the two sister compounds. Due to the random Zn vacancies, quantum oscillations are not  detectable in SrZnBi$_2$. By contrast, pronounced de Haas-van Alphen (dHvA) and Shubnikov-de Haas (SdH) oscillations were observed in BaZnBi$_2$ for the out-of-plane field orientation. Analyzing the dHvA and SdH oscillations we find key signatures of Dirac fermions, including a light effective mass ($\sim 0.1m_e$, where m$_e$ is the free electron mass), high quantum mobility (684 cm$^2$ V$^{-1}$s$^{-1}$), and a non-zero Berry phase accumulated along the cyclotron orbit. Furthermore, we directly probe the Dirac dispersion by angle-resolved photoemission spectroscopy (ARPES) and identify the responsible bands by first-principles density functional theory calculations.

\begin{figure*}[t]
  \includegraphics[width=\textwidth]{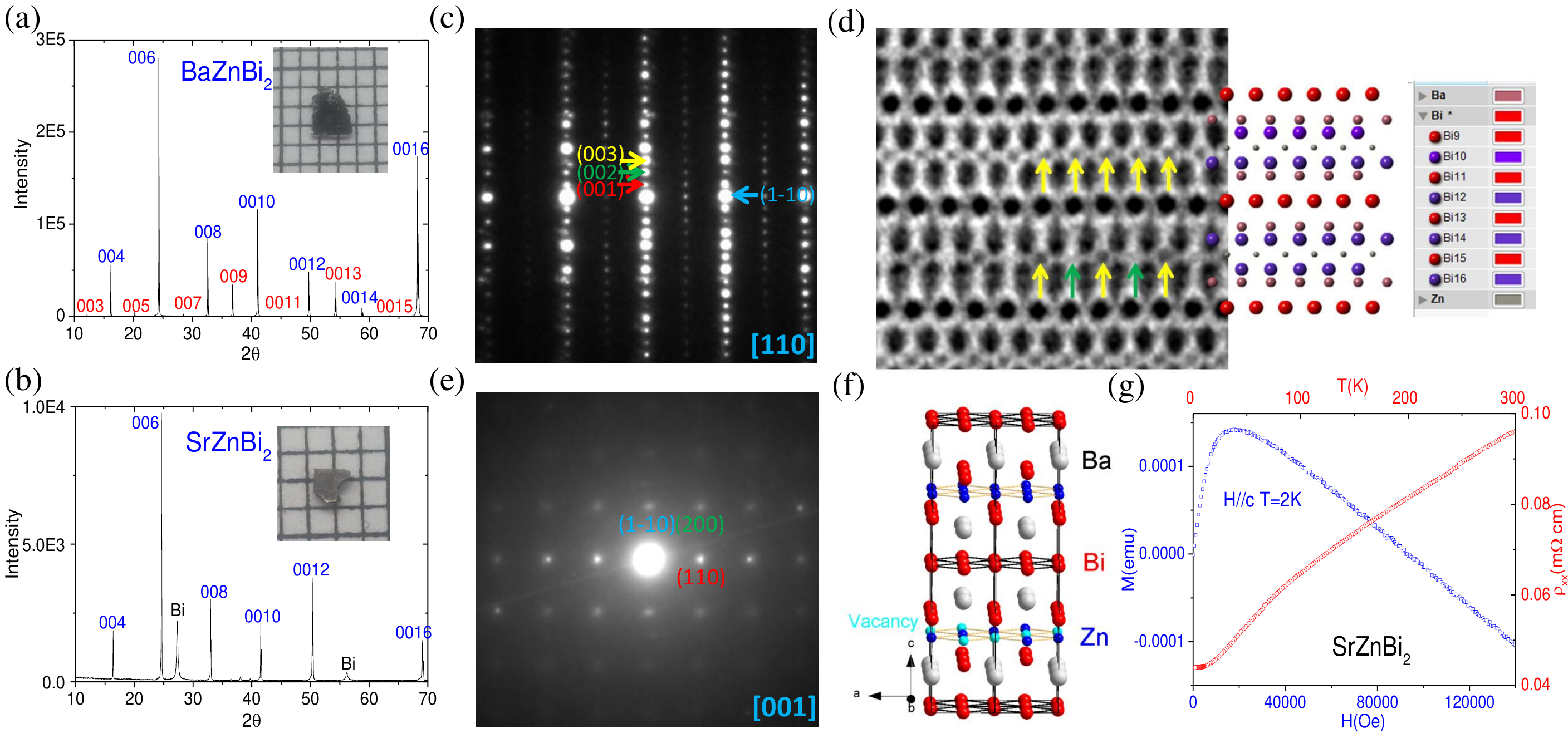}
  \caption{(a)-(b)Single crystal XRD pattern of BaZnBi$_{2}$ and SrZnBi$_{2}$, with an optical image of a typical single crystal as an inset, (c) Selected area electron diffraction (SAED) pattern along [1 1 0] direction, (d) Corresponding atomic resolution scanning transmission electron microscope (STEM) image, with yellow arrow indicating Zn site and green arrow indicating Zn vacancy site, (e) Selected area electron diffraction (SAED) pattern along [0 0 1] direction, (f) Crystal structure of BaZnBi$_{2}$ in the local Zn vacancy order region, with vacancy positions indicated by light blue circles,
  (g) Left: Isothermal out-of-plane (H//c) magnetization for SrZnBi$_2$ at 2K, Right: In-plane resistivity ($\rho$$_{xx}$) of SrZnBi$_2$ between 2K and 300K. (see text).}
  \label{TEM}
\end{figure*}

Fig.\ \ref{TEM}(a) shows the X-ray diffraction pattern of single-crystalline BaZnBi$_2$, with only (00L) reflections observed, in line with the expected preferred orientation of plate-like crystals. SrZnBi$_2$ has a tetragonal structure similar to that of SrMnBi$_2$, as only (0, 0, 2L) peaks are visible in Fig.\ \ref{TEM}(b), consistent with the diffraction pattern of the space group I4/mmm. The lattice parameters of BaZnBi$_2$ agree with those reported in Ref. \cite{BaZnBi2}, $a=b=4.847\AA$, and $c=21.98\AA$. However, Fig.\ \ref{TEM}(a) clearly indicates the presence of (0, 0, 2L+1) peaks as well, signaling a new structural modulation in BaZnBi$_2$, which had not been detected in powder samples.

We have measured selected area electron diffraction (SAED) along [1 1 0] direction in Fig.\ \ref{TEM}(c), on a circle area with the radius of roughly 200~nm. Consistent with X-ray diffraction pattern, (001), (002), and (003) etc., the corresponding SAED reflections are also visible. Weak but clear diffraction peaks appear at (H,-H,0) (H=half integer), indicating the existence of a modulated structure doubling the unit cell along a face diagonal direction, compared to the space group I4/mmm.

Next, we employed atomic resolution scanning transmission electron microscopy (STEM) to directly probe the modulated structure of BaZnBi$_2$ in real space, for a thinned crystal with thickness of about 50~nm. As shown in Fig.\ \ref{TEM}(d), atomic arrangements of Bi and Ba columns are in good agreement with the standard tetragonal structure, while an ordering of Zn vacancies appears in every second Zn layer. Contrary to the occupied sizes, indicated by yellow arrows, the sites marked by green arrows are unoccupied. The [1 1 0] direction is the face diagonal direction of the tetragonal structure, and the Zn vacancies also order along this direction, as illustrated in Fig.\ \ref{TEM}(f).

After the observation of Zn vacancy order in certain local regions, to investigate whether this structural feature forms a long range order, we have measured SAED along [0 0 1] direction on several different circle areas with the radius of roughly 300~nm. In Fig.\ \ref{TEM}(e), the diffraction pattern is consistent with that of the space group I4/mmm in ab plane, with no diffraction peaks at (H,-H,0) (H=half integer), indicating absence of structural modulation in the ab plane of BaZnBi$_2$. In Fig. 1(b), we observe a weak diffraction peak at (1/2, -1/2, 0) in [1 1 0] direction; however, this diffraction peak does not appear in [0 0 1] direction. Thus, combining the SAED pattern along both [1 1 0] and [0 0 1] directions, we conclude that the Zn vacancy ordering has a local character and is not uniform, not forming a long range structure in BaZnBi$_2$. This local Zn vacancy ordering causes large differences on the physical property between BaZnBi$_2$ and SrZnBi$_2$.

There is no indication of quantum oscillations in the isothermal magnetization up to 14 T for SrZnBi$_2$ (at 2K in out-of-plane orientation) in Fig.\ \ref{TEM}(g). In fact, the in-plane residual resistance ratio (RRR) =R(300K)/R(2K)=2.3 is much smaller compared to 8 for SrMnBi$_2$ and CaMnBi$_2$\cite{Sr112PRL, Sr112PRB, Ca112PRB, Ca112APL}, indicating strong scattering, presumably from a large number of disorder Zn vacancies.

\begin{figure}[t]
 \includegraphics[width=\columnwidth]{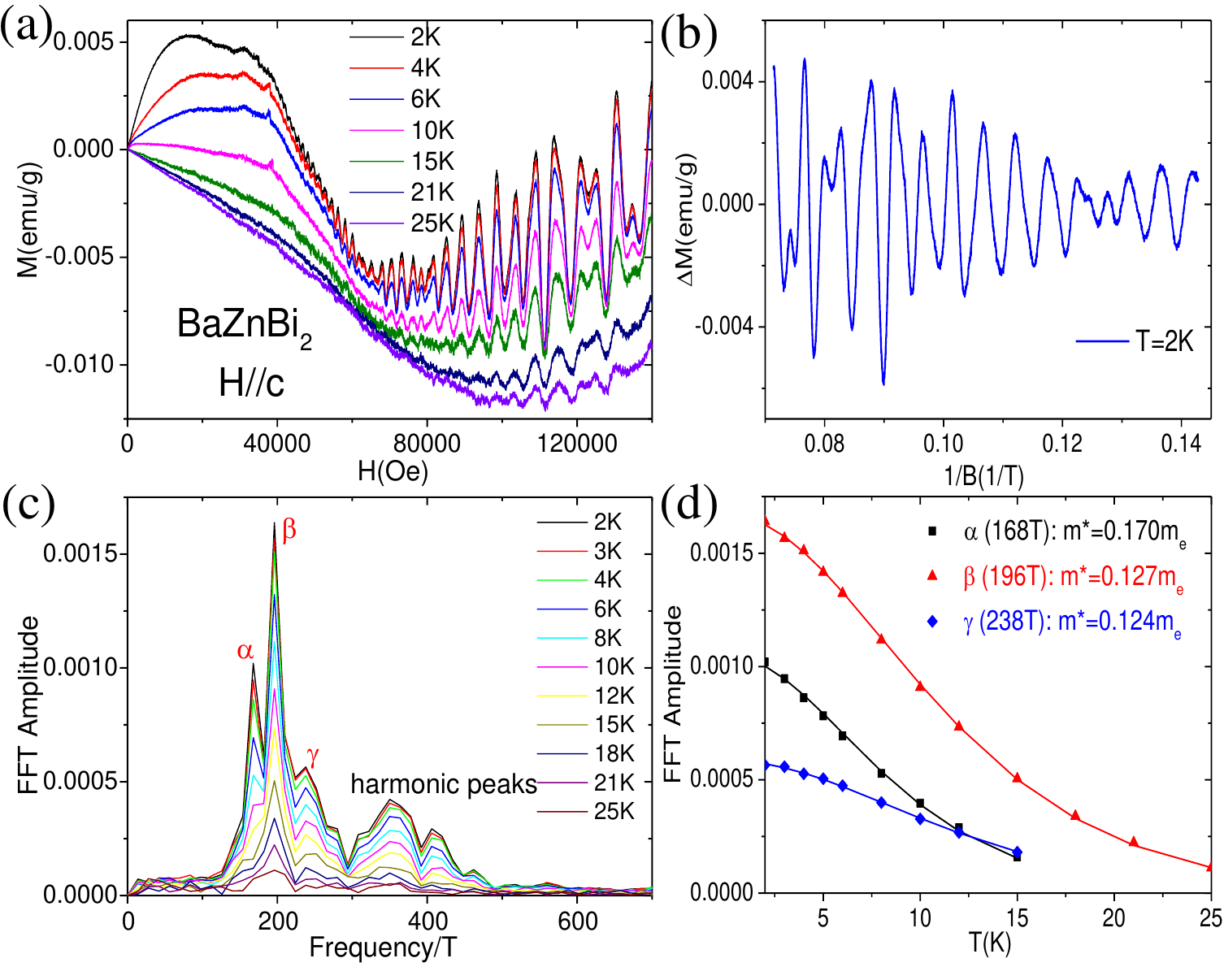}
 \caption{(a) Isothermal out-of-plane (H//c) magnetization for BaZnBi$_2$ at various temperatures, (b) Oscillatory component of magnetization for BaZnBi$_2$ at 2~K between 7~T and 14~T, (c) FFT spectra of $\Delta$M(B) at different temperatures (the FFT was done in the field range between 7 and 14~T),(d) The temperature dependence of the normalized FFT amplitude for $\alpha$, $\beta$, and $\gamma$ oscillations, with fitting to the Lifshitz-Kosevich (LK) formula.(see text).}
 \label{dHvA}
 \end{figure}

Fig.\ \ref{dHvA}(a) shows the isothermal magnetization measured up to 14~T for BaZnBi$_2$ in H//c. The magnetization exhibits strong dHvA oscillations, which start around 5~T and persist up to 25~K, similar as found for CaMnBi$_2$ and SrMnSb$_2$\cite{Ca112APL, SrMnSb2NM}. In Fig.\ \ref{dHvA}(b), we present the oscillatory component of magnetization for BaZnBi$_2$ at 2~K, obtained after subtracting the background between 7 and 14~T. From the fast Fourier transformation (FFT) analysis in Fig.\ \ref{dHvA}(c), three frequencies 168, 196, and 238~T are derived, comparable to the dHvA frequencies 101 and 181~T measured in CaMnBi$_2$\cite{Ca112APL} and the SdH frequency of 152~T in SrMnBi$_2$ \cite{Sr112PRL} for the same field orientation. Evidence for Dirac fermions in BaZnBi$_2$ has been obtained from the further analysis of the dHvA oscillations. As shown in Fig.\ \ref{dHvA}(d), the effective cyclotron mass $m^\ast$ can be extracted from the fit of the temperature dependence of the normalized FFT peak amplitude according to the thermal damping factor in the Lifshitz-Kosevich (LK) equation, i.e. $\Delta$M/M$_0$ $\propto$ AT$\mu$/B/[sinh(AT$\mu$/B)], A=2$\pi$$^2$k$_B$m$_e$/$\hbar$e is the thermal factor, $\mu$ is the ratio of effective mass to the free electron mass. We calculated the FFT within the field range 7$\sim$14~T, with average $B=9.33$~T and obtained effective masses $m^\ast = 0.127m_e$ for the 196~T peak, 0.170m$_e$ for 168~T , and 0.124m$_e$ for the 238~T orbits(see Fig.\ \ref{dHvA}(d)). The effective mass in BaZnBi$_2$ is thus smaller than that in SrMnBi$_2$ (0.29m$_e$ )~\cite{Sr112PRL} and in CaMnBi$_2$ (0.35m$_e$)~\cite{Ca112PRB, Ca112APL}, and comparable to 0.105m$_e$ in BaMnBi$_2$\cite{Ba112PRB} and 0.14m$_e$ in SrMnSb$_2$~\cite{SrMnSb2NM}.

\begin{figure}[t]
 \includegraphics[width=1\columnwidth]{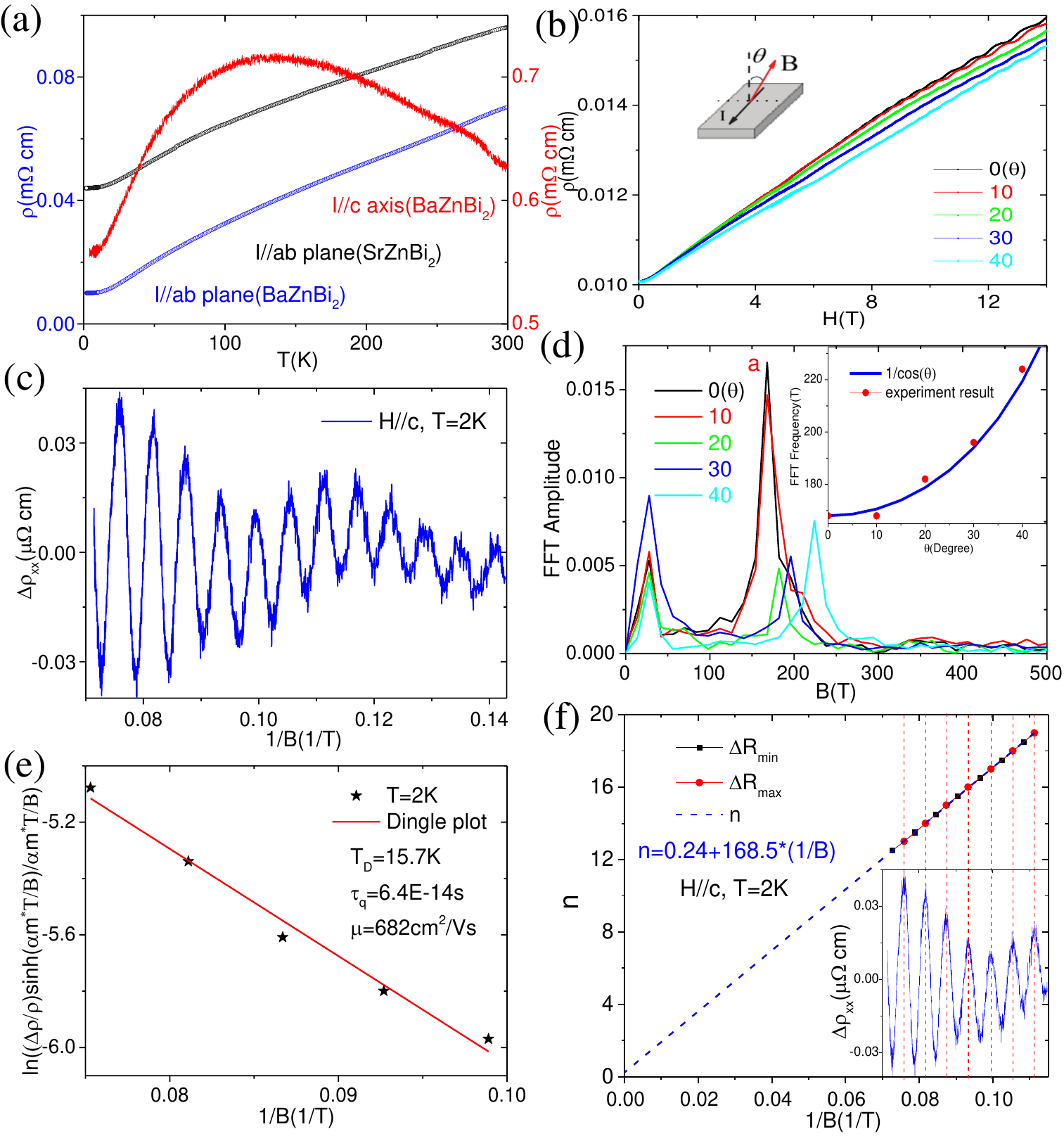}
 \caption{(a) In-plane resistivity ($\rho$$_{xx}$) and out of-plane resistivity ($\rho$$_c$) of BaZnBi$_2$ as a function of temperature, with $\rho$$_{xx}$ of SrZnBi$_2$ as a comparison, (b) Magnetoresistance measured under different field orientations, with the diagram of the measurement setup as inset, $\theta$ is defined as angle between the magnetic field and the out-of-plane direction, (c) the oscillation component of in-plane resistivity, $\rho$$_{xx}$ vs. 1/B measured for H//c at 2K, (d) the FFT spectra of $\Delta$$\rho$$_{xx}$(B) at different field orientations. Inset: the angular dependence of the SdH oscillation frequency. The dashed curve is a fit to F($\theta$) = F(0$^{\circ}$)/cos$\theta$, (e) Dingle plot for the in-plane quantum oscillations $\rho$$_{xx}$ at 2 K, (f) Landau level (LL) fan diagram. The blue dashed line represents the linear fit, with integer LL indices assigned to the maximum of resistivity (see text).}
 \label{SdH}
 \end{figure}

 Due to formation of a local Zn vacancy order in BaZnBi$_2$, the scattering is greatly reduced, and the in-plane resistivity $\rho$$_{xx}$(T) [Fig.\ \ref{SdH}(a)] is metallic with the RRR about 7, comparable to that of (Ca/Sr/Ba)MnBi$_2$\cite{Sr112PRL, Sr112PRB, Ca112PRB, Ca112APL, Ba112PRB}. The out-of-plane resistivity is larger than the in-plane resistivity, with $\rho$$_c$(T)/$\rho$$_{xx}$(T) increasing from less than 10 at 300~K to more than 50 at 2~K.

\begin{figure}[t]
 \includegraphics[width=0.85\columnwidth]{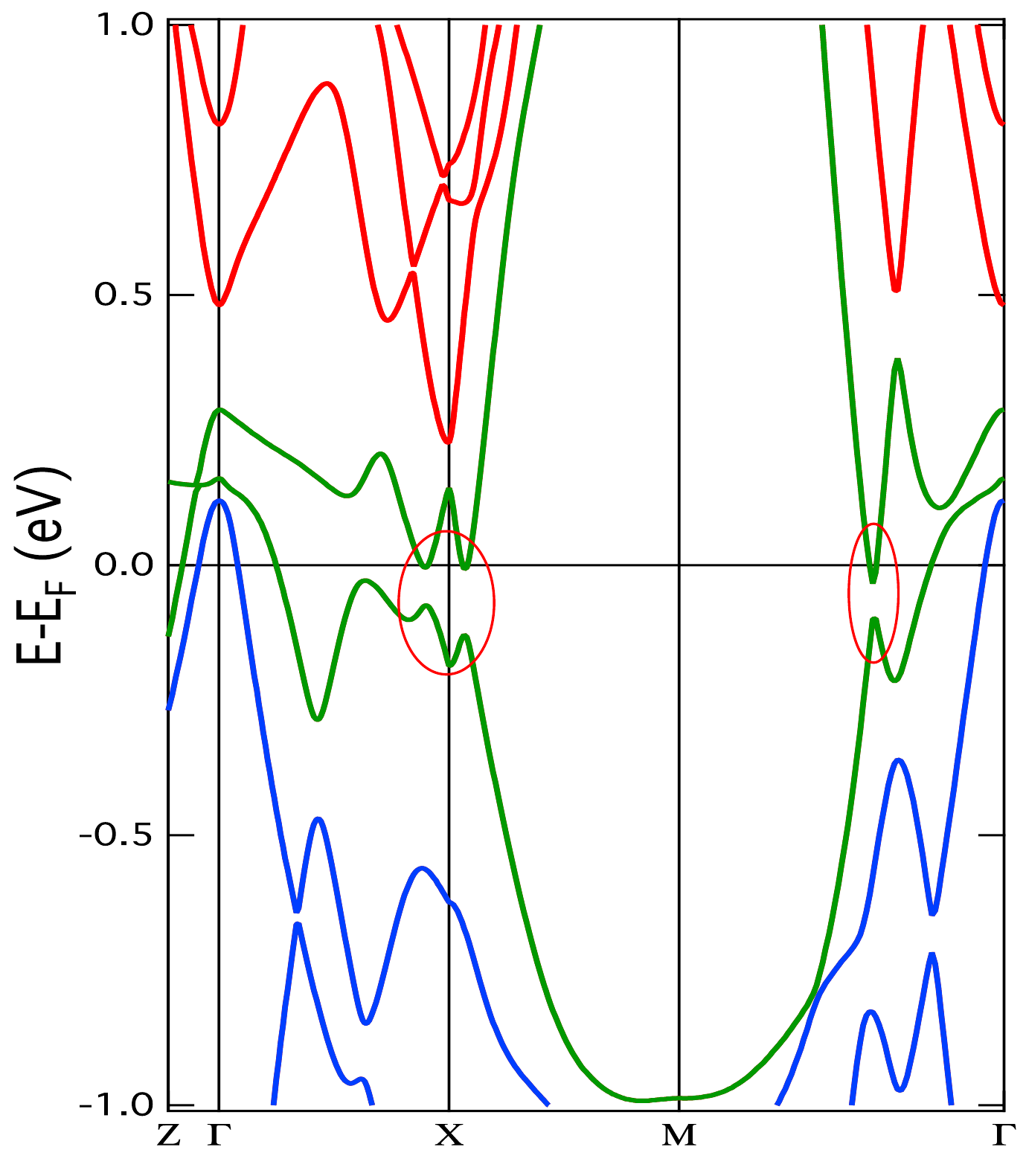}
 \caption{Band structure of BaZnBi$_2$ calculated including SOC with Dirac bands marked by red circles (see text).}
 \label{DFT}
 \end{figure}

\begin{figure*}[th!]
  \includegraphics[width=0.9\textwidth]{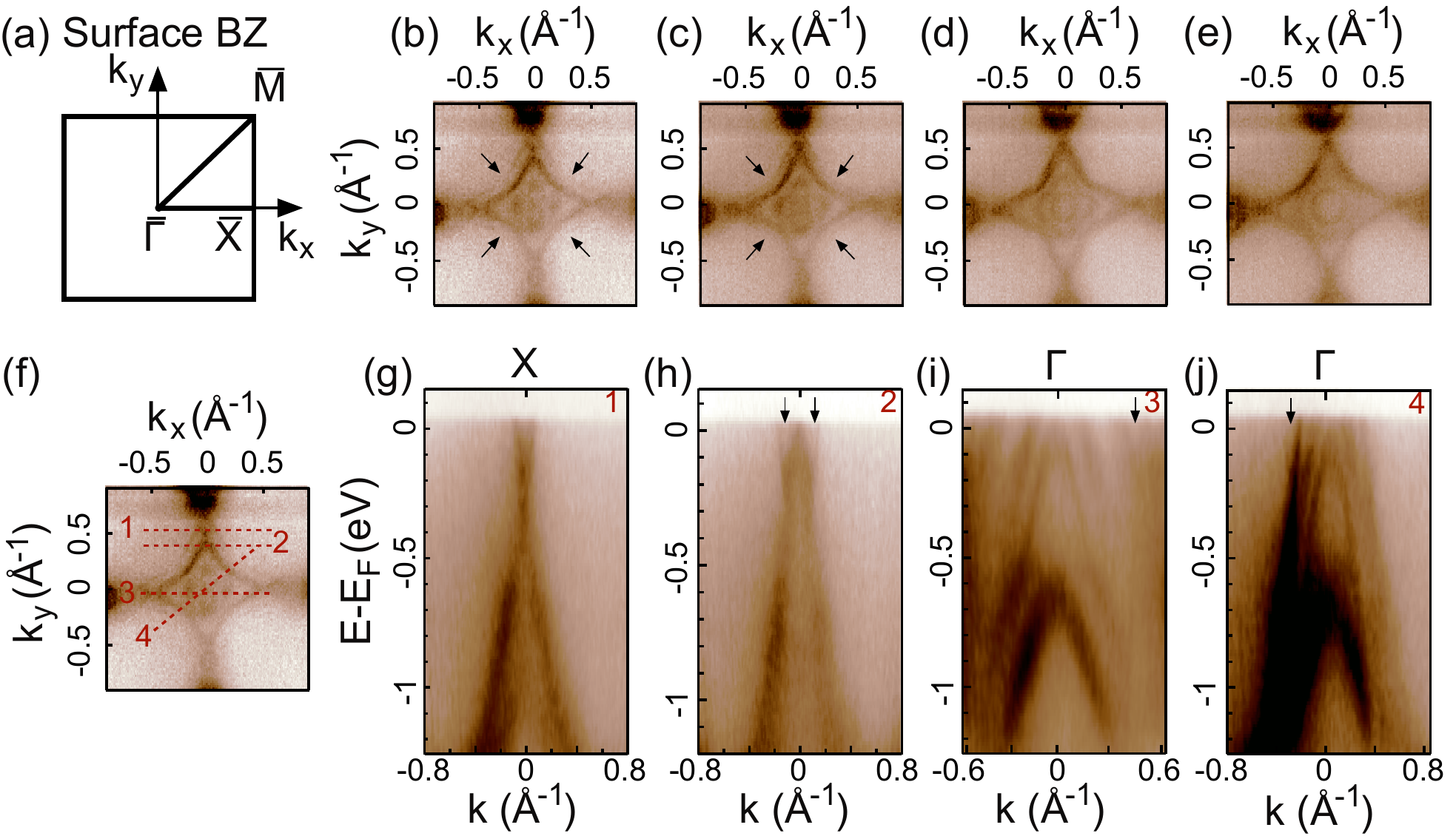}
  \caption{ ARPES results: (a) Sketch of the BaZnBi$_2$ (001) Surface Brillouin Zone corresponding to the high-symmetry directions presented in the constant energy contours of (b)-(e). (b)-(e) Constant energy contour plots at binding energies of 0, 90, 200, and 300 meV, respectively, obtained using horizontally-polarized light with $h \nu = 75$~eV, with black arrows indicating the faint distorted rhombus bands in (b) and (c). (f) Red dashed numbered lines on the Fermi surface of BaZnBi$_2$ highlight the lines of the corresponding band dispersion shown in (g) - (j), with black arrows indicating the crossing of the two distorted rhombus bands seen in (b) and (c).}
  \label{ARPES}
\end{figure*}

Like (Ca/Sr/Ba)Mn(Bi/Sb)$_2$~\cite{Sr112PRL, Sr112PRB, Ca112PRB, Ca112APL, Sr112DFT, Sr112Neutron, Sr112APRES-PRB, Sr112APRES-SR, Eu112SA, BaMnSb2SR, Ba112PRB, Yb112PRB, SrMnSb2NM, Yb112ARPES, BaMnSb2PNAS, CaMnSb2PRB}, BaZnBi$_2$ also exhibits quantum transport behaviors. As shown in Fig.\ \ref{SdH}(b), a systematic evolution of magnetoresistance is observed in a magnetic field rotating from the $c$ axis to the $ab$ plane direction (see the insets of Fig.\ \ref{SdH}(b) regarding the experimental set-up with B//I for $\theta$=90$^\circ$). The SdH oscillatory component is shown in Fig.\ \ref{SdH}(c) between 7 and 14~T with H//c. The oscillation frequency is 168~T, exactly the same as one of the frequencies of the dHvA oscillations. Furthermore, the oscillation frequency F($\theta$) extracted from $\rho$$_{xx}$ can be fitted to F($\theta$) = F(0$^{\circ}$)/cos$\theta$ (inset of Fig.\ \ref{SdH}(d)), suggesting that the Fermi surface responsible for the SdH oscillations is either two dimensional (2D) cylindrical shape or an elongated elliptical shape. Note there is another oscillation with the frequency 30T that does not change with the magnetic field rotation. Similar to other Bi/Sb layer semimetals~\cite{Sr112PRL, Sr112PRB, Ca112PRB, Ca112APL, Sr112DFT, Sr112Neutron, Sr112APRES-PRB, Sr112APRES-SR, Eu112SA, BaMnSb2SR, Ba112PRB, Yb112PRB, SrMnSb2NM, Yb112ARPES, BaMnSb2PNAS, CaMnSb2PRB}, the angular magnetoresistance (AMR) of BaZnBi$_2$ at 2~K in 7~T, shown in supplemental Fig. S1(a) follows a |cos($\theta$)| dependence very well with a typical twofold anisotropy expected for a material with a quasi-2D electronic structure.

Besides the small effective electron mass, Dirac Fermions also have a high quantum mobility. This holds for instance in topological insulators and in Cd$_3$As$_2$\cite{Cd3As22015NM, TIRMP} and is also verified in BaZnBi$_2$. The quantum mobility is related to the quantum relaxation time $\tau$$_q$ by $\mu$$_q= e \tau_q/m^\ast$, which could be found from the field damping of the quantum oscillation amplitude, i.e., $\Delta$$\rho$/$\rho$$_0$ $\propto$ exp(-AT$_D$$\mu$/B)*AT$\mu$/B/[sinh(AT$\mu$/B)]. T$_D$ is the Dingle temperature and is linked to $\tau$$_q$ as T$_D$ = h/(2pkB$\tau$$_q$). With $m^\ast$ already known, $\tau$$_q$ at 2K can be extracted through the linear fit of ln ([B*sinh (AT$\mu$/B)/AT$\mu$]*$\Delta$$\rho$/$\rho$$_0$) against 1/B. As shown in Fig.\ \ref{SdH}(e), we have obtained T$_D$=15.7K, $\tau$$_q$= 6.4$\times$ 10$^{-14}$s, from which the quantum mobility $\mu$$_q$(= e$\tau$$_q$/$m^\ast$) is estimated to be 684 cm$^2$ V$^{-1}$s$^{-1}$, comparable to that of SrMnBi$_2$ (250 cm$^2$ V$^{-1}$s$^{-1}$ ) and SrMnSb$_2$ (~570 cm$^2$ V$^{-1}$s$^{-1}$)\cite{Sr112PRL, SrMnSb2NM}. Compared with the strong dHvA oscillations, the SdH oscillations of BaZnBi$_2$ are relatively weak, probably associated with the strong background magnetoresistance, similar in the nodal line semimetal ZrSiS/Se/Te case\cite{ZrSiSePRL, ZrSiSPRB, ZrGeSPRB}. As discussed before, the scattering has been greatly reduced due to the local vacancy order in BaZnBi$_2$; however, there is still some scattering, especially at the boundary of Zn vacancy order region. 
The 196T and 238T oscillations in magnetization probably arise from bands which are more sensitive to such scattering in the transport experiment. Another possibility is that higher magnetic field is needed to observe them in SdH oscillations. As discussed below, the single dominant SdH frequency actually allows us to extract the respective Berry phase.

To verify the topological nature of the nearly massless electrons in BaZnBi$_2$, we extracted the Berry phase accumulated along the cyclotron orbit from the analysis of SdH oscillations. We present the Landau level (LL) fan diagram constructed from the SdH oscillations of $\rho$$_{xx}$ for BaZnBi$_2$ in Fig.\ \ref{SdH}(f), where the integer LL indices are assigned to the maxima of $\rho$$_{xx}$. The definition of a LL index is based upon a customary practice whereby integer LL indices are assigned to conductivity minima~\cite{TIAndo}. The in-plane conductivity $\sigma$$_{xx}$ can be calculated from the longitudinal resistivity $\rho$$_{xx}$ and the transverse (Hall) resistivity $\rho$$_{xy}$ as $\sigma$$_{xx}$= $\rho$$_{xx}$/($\rho$$_{xx}$$^2$ + $\rho$$_{xy}$$^2$). According to Fig. S1(b)~\cite{Ba112SM}, $\rho$$_{xy}$ is less than 0.1$\rho$$_{xx}$ for B < 14 T, and $\sigma$$_{xx}$$\approx$1/$\rho$$_{xx}$, thus the conductivity minima correspond to the $\rho$$_{xx}$ maxima. As shown in Fig. 3f, the intercept on the LL index axis obtained from the extrapolation of the linear fit in the fan diagram is 0.24, corresponding to a 0.5(1)$\pi$ Berry phase, which is smaller than the expected value of $\pi$ for a 2D Dirac system. As in BaMnBi$_2$\cite{Ba112PRB}, this is likely due to the large contribution of parabolic bands at the Fermi level, which is further verified by the first-principles calculations below. Another possibility is a deviation of the Dirac band from pure linearity close to the Dirac point, arising from a small spin-orbit gap as discussed below. The oscillation frequency derived from the fit is 168.5~T, nearly the same as the frequency obtained from the SdH oscillations, emphasizing the reliability of our linear fit in the fan diagram.

The band structure of BaZnBi$_2$ has been calculated from first-principles using the standard linearized augmented plane wave code WIEN2k\cite{Wien2k}, including spin-orbit coupling (SOC). As indicated in Fig.\ \ref{TEM}, the local Zn vacancy ordering does not form a long-range structure in BaZnBi$_2$. Moreover, the Zn content does not show an obvious decrease, compared with Ba in the energy-dispersive X-ray spectrometer (EDX) measurement. Thus, the composition could be written as BaZn$_{1-x}$Bi$_2$ (x<0.1), for simplicity still using BaZnBi$_2$. With most part of the system being uniform without Zn vacancies and generating the quantum oscillations, we adopt the I4/mmm tetragonal structure without Zn vacancies to calculate the band structure. Similar to SrMnBi$_2$ and CaMnBi$_2$\cite{Sr112APRES-PRB, Sr112APRES-SR}, the low-energy states are formed mainly by the Bi orbitals of the 2D Bi layer. The SOC in Bi, opens a gap between the upper and lower Dirac branches. Formally BaZnBi$_2$ is not a Dirac semimetal, such as Cd$_3$As$_2$~\cite{Cd3As22013PRB, Cd3As22014NM, Cd3As22015NM}, but the SOC gap is small enough for its transport properties to be similar to a real case, so we use ${Dirac}$ ${semimetal}$ ${candidate}$ throughout the paper. According to the calculations (Fig.\ \ref{DFT}), there are two Dirac points near X, and one along the $\Gamma$-M direction, quite similar to the isostructural SrMnBi$_2$ \cite{Sr112APRES-SR}. 2D isoenergetic contours for k$_z$=0 are presented in supplemental Fig. S2, and bear further evidence to the presence of Dirac bands. Indeed, the contour corresponding to E-E$_F$=30 meV shows  Dirac bands in the $\Gamma$X and $\Gamma$M direction, and that at the 35 meV plot contains three Dirac bands in $\Gamma$X, XM, and $\Gamma$M direction, respectively~\cite{Ba112SM}. 

In Fig.\ \ref{ARPES}(b)-(e), we present the ARPES results, including the Fermi surface and constant energy contours of BaZnBi$_2$ at binding energies of 90 and 200 and 300 meV, respectively. In Fig.\ \ref{ARPES}(b), arrows indicate a faint electronic state that rims the Fermi surface. In the vicinity of this faint state there is an intense electronic state in smaller k values and four electron pockets at the X points of the surface Brillouin zone (SBZ). We can also see two hole-like states centered at the $\Gamma$ point of the SBZ. On the other hand, the two outer electronic states with the distorted rhombus shape have an opposite dispersion and merge at a binding energy of ~200 meV. These states can be also seen along different high symmetry directions forming Dirac-like dispersions that cross the Fermi level. More specifically, in Fig.\ \ref{ARPES}(g)-(j) we present the band dispersions along the high-symmetry XM, $\Gamma$X, and $\Gamma$M direction, with black arrows indicating the crossing of the two distorted rhombus bands seen in Fig.\ \ref{ARPES}(b) and (c). The apparent Dirac-like feature around the X point in Fig.\ \ref{ARPES}(g) comes from the anisotropic Dirac states in BaZnBi$_2$, as already observed in SrMnBi$_2$ and CaMnBi$_2$\cite{Sr112PRL, Sr112APRES-PRB, Sr112APRES-SR}. Correspondingly, we observe the band crossing point moving towards the Fermi level in the line dispersion parallel to the XM direction close to the center of the electron pocket in Fig.\ \ref{ARPES}(h), which does not show the expected gap opening from an isotropic Dirac band, such as in graphene, topological insulators, and Cd$_3$As$_2$\cite{Cd3As22014NM, GrapheneNature, TIRMP}. Concerning the band dispersion along $\Gamma$M direction in Fig.\ \ref{ARPES}(j), the band crossing is barely seen due to matrix element effects - the parabolic bands that cross the Fermi level along $\Gamma$M can be better seen if one moves off center (see Fig. S4)\cite{Ba112SM}. As shown in supplemental Fig. S2(d)\cite{Ba112SM}, for the hole-like bands at the $\Gamma$ point, the constant energy contour plot for taken with k$_z$=0.4$\pi$/c matches better the ARPES results presented in Fig.\ \ref{ARPES}(b)-(d), an indication that the excitation energy used in our ARPES measurements corresponds to a k$_z$ that intersects off-centre the three-dimensional Brillouin zone. In principle, our ARPES data is in good agreement with the dHvA oscillations and theoretical calculation results, as we observe three Dirac-like states that contribute to quantum oscillations, namely one along XM direction in Fig.\ \ref{ARPES}(g), another one along $\Gamma$X direction in Fig.\ \ref{ARPES}(i), and third one along $\Gamma$M direction in Fig.\ \ref{ARPES}(j), respectively.

In conclusion, contrary to the general belief that vacancies usually prohibit the detection of intrinsic properties of a system, just like for the SrZnBi$_2$ case, we identify that local Zn vacancy order guarantees the dHvA and SdH oscillations in a non-trivial system BaZnBi$_2$, exhibiting light effective mass ($\sim 0.1m_e$), high quantum mobility (684 cm$^2$ V$^{-1}$s$^{-1}$), and a non-zero Berry phase accumulated along the cyclotron orbit. Further consistent with band-structure calculations the Dirac dispersion is directly observed by ARPES. The new Dirac semimetal candidate BaZnBi$_2$ seems a promising platform to investigate the intrinsic properties of Dirac fermions. Because of the similar quantum oscillations and Dirac dispersion in BaZnBi$_2$ compared to AMn(Bi/Sb)$_2$ (A=Ca, Sr, Ba, Eu, and Yb), our results clearly indicate that the Dirac fermions in the two-dimensional Bi/Sb layers are not influenced by the antiferromagnetism in the MnBi/Sb layer, except for the canted case in YbMnBi$_2$~\cite{Yb112ARPES}.

\acknowledgments
The authors would like to thank Peizhe Tang, Philip Moll, Yurii Skourski, Jin Hu, Hongbin Zhang, Qiang Zou, and Jinguang Cheng for helpful discussions and experimental collaborations. The work was supported by the German Science Foundation through the priority program SPP 1666 and the Emmy Noether program - JE 748/1. Work at IOP was supported by the Strategic Priority Research Program of Chinese Academy of Sciences (Grant No. XDB07030200) and National Natural Science Foundation of China (51522212). I.I.M. acknowledges funding from the Office of Naval Research (ONR) through the Naval Research Laboratory$^{\prime}$s Basic Research Program.

${Note}$ ${added}$: During the preparation and submission of our manuscript, two papers about BaZnBi$_2$ were published\cite{Ba112NJP, BaZn112PRB}, which show the quantum oscillation results. However, with the presence of (0, 0, 2L+1) peaks in single-crystal X-ray diffraction pattern, both papers do not mention the existence of local Zn vacancy order.

\bibliography{Ba112}

\end{document}